\documentclass[apl, amsmath,amssymb, preprint,]{revtex4-1}%
\usepackage{graphicx}
\usepackage{bm}
\usepackage{epstopdf}
\usepackage{sidecap}
\usepackage{amsmath}
\usepackage{amsfonts}
\usepackage{amssymb}%
\providecommand{\U}[1]{\protect\rule{.1in}{.1in}}
\begin{document}
\title{Dynamically tunable Fano resonance in periodically asymmetric graphene nanodisk pair}
\author{Zhengren Zhang$^{1*}$, Xiaopeng Su$^{2}$, Yuancheng  Fan$^{3}$, Pengfei Yin$^{1}$, Liwei Zhang$^{4}$, and Xi Shi$^{2\dag}$}
\address{
$^{1}$ School of Science, Chongqing Jiaotong University, Chongqing 400074, China\\
\email{zhrenzhang@126.com}
$^{2}$ Key Laboratory of Advanced Micro-structure Materials(MOE), Department
of Physics, Tongji University, Shanghai 200092, China\\
$^{3}$ Key Laboratory of Space Applied Physics and Chemistry, Ministry of Education and Department of Applied Physics, School of Science, Northwestern Polytechnical University, Xi'an 710129, China\\
$^{4}$ School of Physics and Chemistry, Henan Polytechnic University, Jiaozuo 454000, China\\
Corresponding author: $^{\dag}$xishi@tongji.edu.cn, $^{*}$zhrenzhang@126.com}

\begin{abstract}
  We present a dynamically frequency tunable Fano resonance planar device composed of periodically asymmetric graphene nanodisk pair for the mid-infrared region. There are two kinds of modes in this structure, that is, the symmetric mode and the antisymmetric mode. The resonance coupling between the symmetric and antisymmetric modes creates a classical Fano resonance. Both of the Fano resonance amplitude and frequency of the structure can be dynamically controlled by varying the Fermi energy of graphene. Resonance transition in the structure is studied to reveal the physical mechanism behind the dynamically tunable Fano resonance. The features of the Fano resonant graphene nanostructures should have promising applications in tunable THz filters, switches, and modulators.
\end{abstract}
\maketitle
Quantum interference in atomic system has led to several fascinating and extraordinary effects \cite{1,2}. One of the most interesting phenomenon is Fano resonance, which originates from the quantum-mechanical interference between a discrete excited state of an atom and a continuum sharing the same energy level \cite{3,4}. Other than the Lorentzian resonance, the Fano resonance possesses a distinctly asymmetric line profile, and it has been found in classical optical systems \cite{5}. Recently, the Fano resonances have also been observed in several plasmonic nanocomplex, and have found its application in switching \cite{6}, slow light \cite{7}, sensing \cite{8}, etc. Due to the interesting physics and abundant applications, generation of physical phenomena equivalent to Fano resonance at terahertz (THz) frequencies is currently one of the most exciting topics of plasmonics research \cite{9,10}. It is known that the noble metal can support surface plasmons (SPs) in THz frequencies \cite{11}. Thus, it is widely used to construct plasmonic nanocomplex which can realized the Fano resonance. To dynamically tunable Fano resonance of the plasmonic system is one key issue in the application of the high quality Fano response. However, the tunability of metal is difficult to achieve. In the normal case, we need to carefully reoptimizing and amplifying or reducing the geometric parameters of the metal structures to realize the tunable Fano resonance. This method is very troublesome and limits the practicality of the Fano resonance for a wide variety of applications.

\begin{figure}[pb]
\centerline{\includegraphics[width=8.3cm]{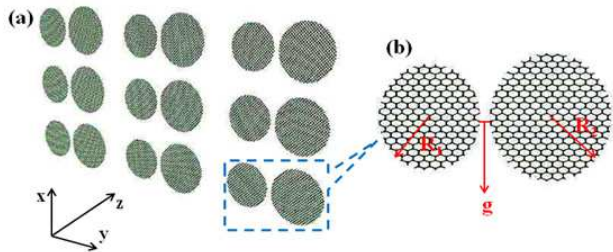}}\caption{(Color
online.) Fig.1. (a) Schematic model of graphene disk nanostructures. (b) A unit cell of structure of our design. Geometric parameters are denoted by red letters.}%
\end{figure}

Graphene, a two dimensional material with only one carbon atom thick, since its exfoliation from graphite \cite{12}, has become a well-known material with unique optical and electrical properties \cite{13,14,15}. As it supports SPs with both high confinement and relatively low loss \cite{13,16,17,18,19}, graphene shows some advantages over metals for plasmonics devices, especially in THz spectral range. Moreover, the dielectric properties of graphene can be dynamically tuned by electrochemical potential via a chemical or electrostatic gating, magnetic field, or optical excitation \cite{20,21,22}. This allows for the creation of surface plasmon based devices that can be effectively turned on and off or tuned to be active at different frequencies. Therefore, it would be a meaningful work to construct graphene nanocomplex to achieve the dynamatically tunable Fano resonance.

 In this paper, we present a controllable periodic structure with its unit cell composed of two asymmetric graphene nanodisks for the excitation of Fano resonance. By studying the local field distributions, we find that two kinds of modes exist in the structure, that is, the symmetric mode and antisymmetric mode. The symmetric mode is a superradiant mode which exhibits a strong radiative damping and is broadened into a continuum. The antisymmetric mode is a subradiant plasmon mode that induces a sharp Fano resonance in the superradiant continuum. Frequency-shift active control of the Fano resonance is realized by varying the Fermi energy of the graphene without reoptimizing and refabricating the nanostructures. Meanwhile, the resonance transition in the structure is demonstrated to well explain the physical mechanism of the dynamically tunable Fano resonance.

 The schematic of the asymmetric graphene disk nanostructure is presented in Fig.1 (the graphene sheet has the thickness of $d$=1nm). The radiuses of the two graphene nanodisks in the unit cell are $R_{1}$ and $R_{2}$, respectively. The two asymmetry graphene nanodisks are on the same plane, and the distance between them is denoted by $g$ ($g$=68nm). We suppose that the structure is surrounded by air for simplicity and illuminated along the z direction. The designed devices are investigated in the mid-infrared frequency range at around 25$\sim$40 THz.

 All the designed graphene-based planar structures in this paper are numerically studied by utilizing the commercial package CST Microwave Studio. The conductivity of graphene $\sigma$ is computed within the local random phase approximation limit at the zero temperature [23], which can be written as\\
 $
\sigma(\omega)=\frac{ie^{2}E_{F}}{\pi\hbar(\omega+i\tau^{-1})}+\frac{e^{2}}{4\hbar}[\theta(\hbar\omega-2E_{F})+\frac{i}{\pi}|\frac{\hbar\omega-2E_{F}}{\hbar\omega+2E_{F}}|]
$\\
$\tau=\mu E_{F}/e\upsilon _{F}^{2}$ is the relaxation rate with $\mu=10^{4}cm^{2}/Vs$ and $\upsilon_{F}\approx10^{6}m/s$, the mobility and Fermi velocity, respectively. $E_{F}=\hbar\upsilon_{F}\sqrt{\pi|n|}$ can be easily controlled by electrostatic doping via tuning charge-carrier density $n$. In the mid-IR frequency range, such a conductivity form indicates strong plasmonic response. As can be seen from the equation, changing the Fermi energy enables the control of the propagation characteristics of the graphene plasmons.

\begin{figure}[ptb]
\centerline{\includegraphics[width=8.3cm]{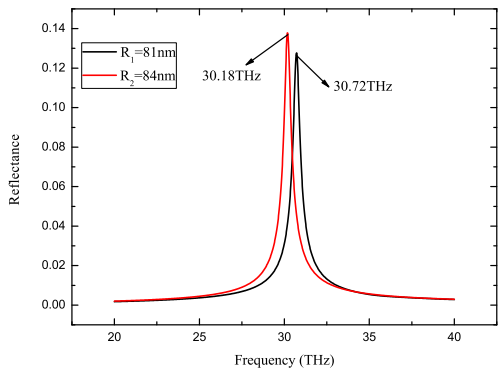}}\caption{(Color
online.) The reflectance spectra of graphene nanostructures with only one nanodisk ($R_{1}$=81nm or $R_{2}$=84nm in radius, the Feimi energy is set to be $E_{F}$=0.5eV) in the unit cell.}%
\end{figure}

To begin with, the reflectance spectra of graphene nanostructures with only one nanodisk (81nm or 84nm in radius, the Feimi energy is set to be $E_{F}$=0.5eV) in the unit cell are presented in Fig.2. For the small graphene disk periodic nanostructure, there is a reflected peak in its reflection spectrum at $f$=30.72THz. For the big graphene disk periodic nanostructure, the reflected peak appears at $f$=30.18THz. The reflected peaks for the two kinds of structures in the reflection spectrum are caused by the exciting of SPs resonance in graphene nanodisks. It is known that the phase-advance oscillation appears below the resonant frequency and delayed phase oscillation occurs above the resonant frequency in a resonance system. From the Fig.2 we can find that the small and big graphene disk periodic nanostructure exhibit different behavior for the external field response at the range of 30.18$\sim$30.72 THz and same behavior at other frequency range. Thus, a combination of the two structures will lead to the occurrence of constructive and destructive interference, which can produce Fano resonance.

Subsequently, we investigate the periodically asymmetric graphene nanodisk pair strucure which has two graphene nanodisks in the unit cell (see Fig.1(a)). As we expected, there is an asymmetric Fano line-shape appears in the reflectance spectrum, as shown in Fig.3 (a). How do we to understand the origin of the Fano resonance nature on this system? In order to reveal the mechanism behind the Fano resonance of this structure, Fig.3 (b)-(d) show the field distributions in the disks' cross section at certain frequencies. The corresponding frequencies are also marked in Figs. 3(a). From the Fig.3 (b)-(d) we can find that the field distributions exhibit a symmetric mode at $f$=29.74 THz, 31.12THz (we find that the field distribution in I and III are also symmetric mode), and exhibit a antisymmetric mode at $f$=30.18 THz (we find that the field distribution in II is also antisymmetric mode). In the symmetric mode, the two disks support in-phase oscillations, and it leads to the constructive interference happen between the reflected light comes from the two kinds of graphene disks. The symmetric mode provides a much broader superradiant dipole mode (strongly coupled to free space), so it can be regard as a continuum state approximation (see I and III in Fig.3). However, in the antisymmetric mode, the two disks support out of phase oscillations. Duo to this reason, the destructive interference occurs between the reflected light comes from the two kinds of graphene disks. The antisymmetric mode occurs in a very narrow frequency range provides a subradiant mode (weekly coupled to free space), and thus, it can be regard as a discrete state approximation (see II in Fig.3). Therefore, the asymmetric Fano resonance in this system arises from the structural asymmetry which leads to an interference between a sharp discrete resonance and a much broader continuum-like spectrum of dipole resonance.

\begin{figure}[ptb]
\centerline{\includegraphics[width=8.3cm]{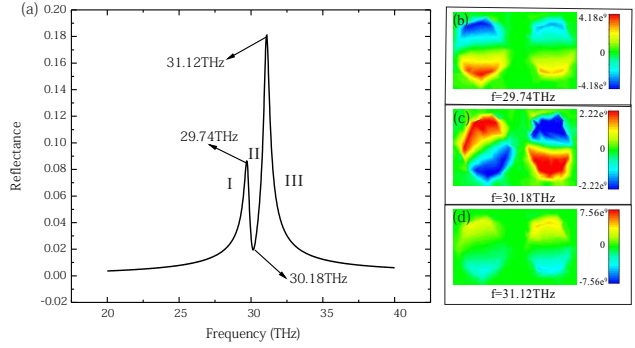}}\caption{(Color
online.) (a) The reflection spectrum of the periodically asymmetric graphene nanodisk pair structure (see Fig.1). Parameters are $R_{1}$=81nm $R_{2}$=84nm and $E_{F}$=0.5eV.
(b) The field distributions in the disks' cross section of the nanostructure at frequencies 29.74, 30.18 and 31.12THz.}%
\end{figure}

 Furthermore and most importantly, one of the major advantages of the designed structures with graphene is the tuneability \cite{24}. This property allows them to work at different frequencies without reoptimizing or reconstructing the physical structure. This is highly desirable in many practical applications since it is very difficult to change the physical structure after fabrication. To demonstrate the frequency tunability of a Fano resonance using the proposed asymmetric graphene disk nanostructures, we graph the reflectance of our structure under different Fermi energy, as shown in Fig.4. In this simulation, Fermi energy varies 0.5eV to 0.6eV and the other parameters are same as in Fig.3. From the Fig.4 we find that the resonant strength of the two kinds of graphene disks is enhanced and blue shifted as the increase of Fermi energy. This behavior can be interpreted through the resonance condition. The wave vector of surface plasmons along the graphene disk satisfies $k_{SP}\propto\frac{1}{D}$, and $k_{SP}=\frac{\hbar \omega _{r}^{2}}{2\alpha E_{F}c}$, where $D$ represents the diameter of the graphene disk, $\alpha=e^{2}/\hbar c$ is  the  fine structure constant, and the $\omega_{r}$ is  the resonance frequency \cite{23}. Therefore, the resonance frequency can be described as $\omega _{r}\propto \sqrt{2\alpha E_{F}c/\hbar D}$. Hence, the resonance and Fano frequencies can be tuned by varying the Fermi energy while fixing the geometrical parameters, indicating that graphene nanostructures are more active for the Fano effect than metallic ones, and thus may find applications in THz sensing and communications.

\begin{figure}[ptb]
\centerline{\includegraphics[width=8.3cm]{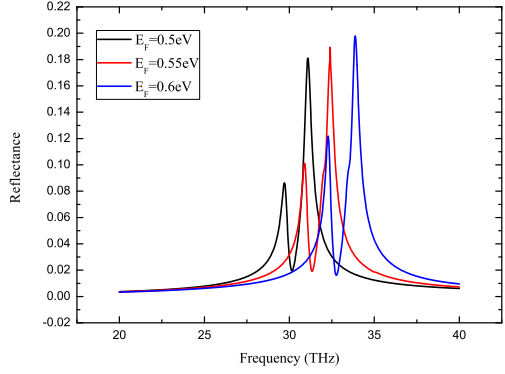}}\caption{(Color
online.) The reflection spectrum of the periodically asymmetric graphene nanodisk pair structure (see Fig.1) for different $E_{F}$. The other parameters are same as in Fig.3.}%
\end{figure}

  In summary, we have proposed a dynamically tunable Fano planar nanostructure for the mid-infrared region based on graphene disks. The Fano resonance can be realized using a single layer of graphene disks for a fixed Fermi energy. The resonance results from the interaction between the reflected light comes from the two different kinds of graphene disks in our structure. In addition, by varying the Fermi energy of the graphene, the frequency of the Fano resonance can be dynamically tuned without the need to reoptimize or refabricate the nanostructures. This makes graphene Fano device more useful than metallic Fano devices. We believe that this method of actively tuning the frequency of Fano may open up avenues for the development of compact elements such as tunable sensors, switchers and so on.

This work is supported by Science Foundation of Chongqing Jiaotong University (Grant No. 2013kjc030), the Fundamental Research Funds for the Central Universities (Grant No. 3102015ZY079), and the Foundations of Henan Educational Committee (Grants No. 14A140011 and 2012GGJS-060).

\end{document}